\documentstyle[12pt,epsfig]{article}
\newcommand{\be}{\begin{equation}}
\newcommand{\ee}{\end{equation}}
\newcommand{\Tr}{\mathop{\rm Tr}\nolimits}

\begin{document}
\title{
\begin{flushright}
{\small SMI-5-98 }
\end{flushright}
\vspace{2cm}
Chaos-Order Transition in Matrix Theory}

\author{
I. Ya. Aref'eva${}^{\S}$, A.S. Koshelev and P. B. Medvedev${}^{\star}$\\
\\${}^{\S}$ {\it  Steklov Mathematical Institute,}\\ {\it Gubkin st.8, Moscow,
Russia, 117966}\\
arefeva@genesis.mi.ras.ru\\
\\${}^{\star}$
{\it Institute of Theoretical and Experimental Physics,}\\
{\it B.Cheremushkinskaya st.25, Moscow, 117218}\\
medvedev@heron.itep.ru\\
\\${}^{\dag}$
{\it Physical Department, Moscow State University, }\\
{\it Moscow, Russia, 119899} \\
kas@depni.npi.msu.su
}

\date {$~$}
\maketitle
\begin {abstract}
Classical  dynamics in $SU(2)$ Matrix theory is investigated.
A classical chaos-order transition is found.
For the angular momentum small enough (even for small coupling constant)
the system exhibits a chaotic behavior, for angular momentum large enough
the system is regular.
\end {abstract}
\newpage
\section{Introduction}
Matrix theory
\cite {BFSS} is a surprisingly simple quantum mechanical model
that is able to describe some major properties of superstring theory.
Therefore the model obviously deserves a thorough study. Calculation
of physical quantities is reduced to the appropriate
calculations in the matrix quantum mechanics.  A system of $N$ Dirichlet
zero-branes is described in terms of nine $N\times N$ Hermitian matrices
$X_i, i=1,...,9$ together with their fermionic superpartners. The action can
be regarded as the dimensional reduction of
ten-dimensional $SU(N)$ supersymmetric Yang-Mills to
$(0+1)$ space-time dimensions:
\begin{equation}
\label {1.1}
S=\int dt Tr(\frac{1}{2}D_tX_iD_tX_i+\frac{1}{4}
[X_i,X_j][X_i,X_j] )+ (fermions),
\end{equation}
where $D_t=\partial_t +iA_0$.

The action  (\ref {1.1}) was considered in
the theory of eleven-dimensional
supermembranes in \cite {WHN,WLN,FH} and in the dynamics of D-particles
in  \cite {Wit,DFS,KP}. In the original formulation \cite {BFSS}
of the conjectured correspondence between M-theory and M(atrix) theory
the large $N$ limit was assumed. A more recent formulation
\cite {SUS} deals with finite $N$.
In the last year the Matrix  theory
was a subject of numerous investigations, for reviews see for
example \cite {Ban,SUS,Tay}.

Although the model (\ref {1.1}) is relatively simple its classical dynamics
is in fact rather complicated.
In this note we discuss the bosonic sector of (\ref {1.1})
following the lines of our previous paper \cite {AMRV}.
There we have realized that at least at some specials cases, the solutions
of classical equations of motion
\begin {equation} \label {1.2}
\ddot{X}_i =[[X_j,X_i],X_j]
\end{equation}
are exponentially unstable, i.e. the system (\ref{1.2}) is stochastic.
The
appearance of chaos in a classical system means that we cannot trust to the
ordinary semiclassical analysis of the corresponding quantum system.
Let us note that  $\alpha^\prime$ corrections to the action
(\ref{1.1}) induce a stabilization  of the classical trajectories \cite{AFK}.

Here we shall confine ourselves to the simplest version of (\ref{1.1})
which corresponds to the reduction of
(2+1) dimensional $SU(2)$ Yang-Mills to (0+1). In the $A_0=0$
gauge we deal with eqs.(\ref{1.2}) for $i=1,2$ and the Gauss law constraint
\begin {equation} \label {1.3}
[X_i,\dot{X}_i]=0.
\end{equation}

High symmetry of the system allows one to reduce the dimension of the
phase space. Three components of the Gauss law and one more first
integral, which we denote by $n$, lead to a four dimensional phase space
and the Hamiltonian
\be
H=\frac12 (p_f^2+p_g^2)+\frac12\lambda f^2g^2
+\frac{n^2}{4(f-g)^2} +\frac{n^2}{4(f+g)^2},
\label{h1}
\ee
where $p_{f,g}$ are the momenta conjugated to the $f,g$.

In this paper we present analytical and numerical study of the system
(\ref{h1}). We will show that for $n$ large enough the system is integrable
and its motion is located in a compact region of configuration space.
For
$n$ small enough the system exhibits the chaotic behavior.
Chaotic behavior is a typical feature of systems which one gets as a
long-wave approximation in a field theory, see for example \cite{BMS,CS,Shur,CGM}.

For a better understanding of a role of $n$-dependent terms in
(\ref{h1}) we start with the toy model governed by the Hamiltonian $H$
with $n=0$ and an infinite elastic reflecting wall parallel to the
$g$-axis.  This model exhibits a chaos-order transition.  For the wall far
enough from the origin the motion is confined to the region $g\ll f$ where
it admits an analytical investigation \cite{Med}.  We show that this model is
integrable in this region.

As the next approximation (especially valid for $g\ll f$ ) we choose the slightly
simplified version of (\ref{h1}) with the hyperbolic
wall potential:
\be
V=\frac{\lambda}{2}f^2g^2 +\frac{n^2}{2f^2}.
\ee
We show numerically that it also exhibits chaos-order transition governed
by the parameter $\zeta
=\frac{4\sqrt{2}}{3\sqrt{3}}\frac{E^{3/2}}{n^2\sqrt{\lambda}}$ and give
some analytical arguments in favor of this result.
Note that an effect of the extra term
$1/2f^2$ to chaotic behavior of a two-dimensional system has
been discussed in the recent paper \cite{CGM}.

All this reasoning forces us to conjecture that Hamiltonian (\ref{h1}) also
describes two phases depending on the value of $n$. We compute the
Poincare sections for a number of characteristic values of $n$ with the energy $E$
and $\lambda$ being fixed. As it expected for small $n$ one has the
typical stochastic distribution of points and for large $n$ the points
are distributed along regular lines.

The paper organized as follows:
In Section 2 we present out notation and remind the results of
\cite{AMRV}. Section 3 is devoted to a toy model with an
elastic  reflecting  wall.  In Section 4 we discuss the model with hyperbolic
wall  and in Section 5 we present the results of numerical calculations.
\section{Notations}
\renewcommand{\theequation}{\thesection.\arabic{equation}}
\setcounter{equation}{0}
In this section we review the results of \cite{AMRV} concerning the
appropriate parametrization of the configuration space.

The Lagrangian admits two global continuous symmetries.
The $SU(2)$ rotations
\be
X_1^{'}=U^{+}X_1 U,\qquad X_2^{'}=U^{+}X_2 U,
~~~U\in SU(2)
\ee
yields the  conservation
of the "angular momentum"
\be
\label{M}
M =[X_1,\dot X_1]+[X_2,\dot X_2].
\ee
$M=0$ is the Gauss law.
The $O(2)$ subgroup of Lorenz $SO(1,2)$
\begin{eqnarray}
X_1^{'}&=&\cos\varphi\, X_1-\sin\varphi\, X_2,\nonumber \\
X_2^{'}&=&\sin\varphi\, X_1+\cos\varphi\, X_2,
\label{n}
\end{eqnarray}
give one more first integral
\be
N=\Tr(X_1\dot X_2-\dot X_1X_2).
\ee

Hence, it is convenient to parametrize $X_1$ and $X_2$ as follows:
$$
X_1(t)=\sqrt{2}U^{+}(t)\left(\frac{\sigma_3}{2}f(t)\cos\theta(t)-
\frac{\sigma_2}{2}g(t)\sin\theta(t)\right)U(t),
$$
\be
X_2(t)=\sqrt{2}U^{+}(t)\left(\frac{\sigma_3}{2}f(t)\sin\theta(t)+
\frac{\sigma_2}{2}g(t)\cos\theta(t)\right)U(t),
\label{uu}
\ee
where $f(t),\;g(t),\;\theta(t)$ are real functions and $U(t)$ is a
SU(2) group element.

The parametrization (\ref{uu})can be justified as follows.
The variables $X_1$ and $X_2$ could be treated as vectors
in the internal isotopic space.
At any time they can be rotated to belong to some coordinate, say (2,3) plane
by using an $U(t)\in  SU(2)$:
\be
X_1=(0,\,X_1^2,\,X_1^3),\qquad X_2=(0,\,X_2^2,\,X_2^3),
\label{p}
\ee
that fixes $U(t)$ up to rotation around the 1-axis.
This rotation could be used to impose the following constraint:
\be
X_1^2 X_1^3 +X_2^2X_2^3=0.
\label{co}
\ee
The rotation angle to fulfill (\ref{co}) is
$$
\tan 2\chi=-\frac{2(X_1^2 X_1^3 +X_2^2X_2^3)}
{(X_1^2)^2 +(X_2^2)^2 -(X_1^3)^2 -(X_2^3)^2}.
$$

The constraint (\ref{co}) has a transparent geometrical meaning. Let
$\Phi^2 ,\, \Phi^3$ be the following two-vectors: $\Phi^2 =(X_1^2,X_2^2)$
and $\Phi^3 =(X_1^3,X_2^3)$, then (\ref{co}) is the orthogonality
condition: $(\Phi^2,\Phi^3)=0$. A pair of orthogonal vectors in the plane
can be parametrized by two radii and one angle (phase), say:
\begin{eqnarray}
\Phi^3&=&f(\cos\theta ,\sin\theta)\nonumber \\
\Phi^2&=&g(-\sin\theta ,\cos\theta ).
\label{p1}
\end{eqnarray}
Eqs. (\ref{p1}) plus $SU(2)$ rotation give the parametrization (\ref{uu}).
Note, that $U(1)$ angular momentum $N$ just generates the shifts in $\theta$.

The main advantage of the coordinate system described above is that
four of the six Lagrangian equations of motion appear to be nothing but
the Noether conservation laws
\be
\dot M =0~~~\mbox{and}~~~\dot N =0.
\label{cl}
\ee
Taking into account the Gauss law one gets from (\ref{cl}):
\be\label{ecl}
l_1 =\frac{2nfg}{(f^2 -g^2)^2 },~~~ l_2 =0,~~~l_3 =0,~~~
\dot\theta =\frac{n(f^2+g^2)}{(f^2 -g^2)^2},
\ee
where $\dot U U^+ =l=\frac i2 \sigma_j l_j$ and $n=N$.
By substituting (\ref{ecl}) into  Lagrangian equations for $f$ and $g$ one
gets:
\begin{eqnarray}
\ddot f&=& n\frac{f(f^2 +3g^2)}{(f^2 -g^2)} -\lambda fg^2 \label{f}\\
\ddot g&=& n\frac{g(g^2 +3f^2)}{(f^2 -g^2)} -\lambda f^2 g\label{g}
\end{eqnarray}
It is a matter of simple algebra to prove
that eqs.(\ref{f}) and (\ref{g}) are the equations of motion following
from the Lagrangian
\be
L=\frac12 (\dot f^2 +\dot g^2) -
\frac {\lambda}{2} f^2g^2-\frac{n^2}{4(g-f)^2}-\frac{n^2}{4(g+f)^2}.
\label{L}\ee
Just this Lagrangian will be the subject of our subsequent analysis.
In the particular case $n=0$ the Lagrangian (\ref{L}) was the object of intensive
study about fifteen years ago in the context of long-wave approximation of
Yang-Mills theory, this model will be referred as the {\em hyperbolyc
model}. The dynamical system (\ref{L}) has an additional potential term which
produce two reflecting walls along the $f=g$ and $f=-g$ axes. The appearance of the
reflecting walls can crucially change the behavior of the system. To demonstrate
this in the next section we start with study of the hyperbolic model with elastic
reflecting wall.
\section{Hyperbolic model with reflecting wall.}
It is well known that the hyperbolic model exhibits a chaotic behavior
\cite {BMS,CS}.
Equations of motion for this system have the form:
\begin{eqnarray}
\ddot f&=& -\lambda fg^2\label{fw}\\
\ddot g&=& -\lambda f^2 g\label{gw}
\end{eqnarray}
As it is shown in \cite{Med}
in the asymptotical regime, where $y \ll x$ one can integrate the system of
equations (\ref{fw}) and (\ref{gw})
by using the Bogolyubov-Krylov method \cite {Bog}:
\begin{eqnarray}
f(t)&=& -\alpha\frac{t^2}2+\beta t+\gamma \label{ff}\\
g(t)&=& \sqrt{\frac{2\alpha}{\lambda (-\alpha\frac{t^2}2+\beta t+\gamma )}}
        \cos \sqrt{\lambda}[-\alpha\frac{t^3}6+\beta\frac{t^2}2+\gamma t+\varphi_0]
\label{gg}
\end{eqnarray}
This solution is characterized by four parameters $\alpha,\beta,\gamma$ and $\varphi_0$,
which are related with Couchy's initial data
$f_0\equiv f(0),~p_0\equiv \dot f(0)~g_0\equiv g(0)~q_0\equiv \dot g(0)$
as:
\be
f_0= \gamma,~
p_0=\beta,~
g_0= \sqrt{\frac{2\alpha}{\gamma\lambda}}\cos \varphi_0,~
q_0= -\sqrt{\frac{2\alpha\gamma}{\lambda}}\sin \varphi_0
\ee
or more explicitly,
\be
\alpha=\frac{q^2_0+\lambda f_0^2g_0^2}{2f_0},~
\varphi_0 =\arccos \sqrt{\frac{f_0^2g_0^2}{q^2_0+\lambda f_0^2g_0^2}}
\ee
Note that the parameter $\alpha$ is the Ehrenfest adiabatic invariant
\be
\alpha = \frac{\dot f^2+\lambda f^2g^2}{2f}.
\ee
In the region $f>0$, $\alpha$ is positive and therefore there exists
a maximum of coordinate $f$:
\be
f_{max}=\gamma+\frac{\beta^2}{2\alpha}
\ee
$f_{max}$ being expressed in terms of dynamical variables
is an integral of motion
\be
f_{max}=\frac{(\dot f^2+\dot g^2+\lambda f^2g^2)}{(\dot g^2+\lambda f^2g^2)}.
\ee
Energy $E=\frac 12(\dot f^2+\dot g^2)+\frac{\lambda}2f^2g^2$ is related to $f_{max}$
as
\be
E=\alpha f_{max},~{\rm or}~
f_{max}=\frac{2Ef}{\dot g^2+\lambda f^2g^2}
\ee
Note that $\alpha$ and $f_{max}$ are approximate integrals of motion.

In the asymptotic region $g\sim\xi f$ where $\xi \ll 1$, one has
$\frac {\dot f^2}{f^2} \sim \xi \sqrt{2E\lambda} $ and $\frac{\dot f^2}{f^4}
\sim \xi^2 \lambda$.

Let us  put an elastic reflecting wall at $f=l$.
This means that we consider equation (\ref{fw}) and (\ref{gw}) only for $f\geq l$, and $g$ is
an arbitrary. We assume that $f$-component of momenta changes a sign upon
collision with a wall and $g$ component does not change it.
For the case of elastic reflecting wall located on $f=l$ there is maximal
allowed value of $g$,
$g_{max}=\frac 1l\sqrt{\frac{2E}{\lambda}}\ll 1$ .
The characteristic parameter $\xi$ in this case is $\xi\leq\frac{1}{l^2}
\sqrt\frac{2E}{\lambda}$. If one consider a trajectory starting from
a point on the right of the wall with $\xi \ll 1$, then the trajectory is
described rather well by (\ref{ff}) and (\ref{gg}).
After reflecting the particle moves along the trajectory
still given be (\ref{ff}) and (\ref{gg})
with new initial data.
It is evident
that the energy and the Ehrenfest invariant are conserved upon
a collision of the
particle with the elastic wall, so the value of maximal deviation is also conserved.
Therefore, the particle can never reach $f=\infty$. This is basic property
of the hyperbolic model with reflecting wall located so that the
characteristic parameter $\xi$ is small enough. In other words,
we conclude that if we put the reflecting wall so that $\xi \ll
1$ then
we deal with the integrable system and $f_{max}$ is one of its integral
of motion.
\section{Model with Hyperbolic Potential.}
In the asymptotic region $g\ll f$ Lagrangian (\ref{L}) has form:
\be
L=\frac 12(\dot f^2+\dot g^2) -\frac {\lambda}2f^2g^2-\frac{n^2}{2f^2}
\label{lmodel}
\ee
as compare with (\ref{L}) we neglect the  $o(\frac{g^2}{f^2})$ terms.
Equations of motion for the Lagrangian (\ref{lmodel}) are:
\begin{eqnarray}
\ddot f&=&-\lambda g^2f+\frac{n^2}{f^3}\label{fm}\\
\ddot g&=&-\lambda f^2g\label{gm}
\end{eqnarray}
We call this model the hyperbolic model with the hyperbolic wall.
The equipotential lines for (\ref{lmodel}) are defined by:
\be
E=\frac {\lambda}2f^2g^2-\frac{n^2}{2f^2}
\ee
or explicitly:
\be
g=\frac 1{f^2}\sqrt{\frac{2Ef^2-n^2}{\lambda}}
\label{contour}
\ee

\begin{figure}
\epsfig{file=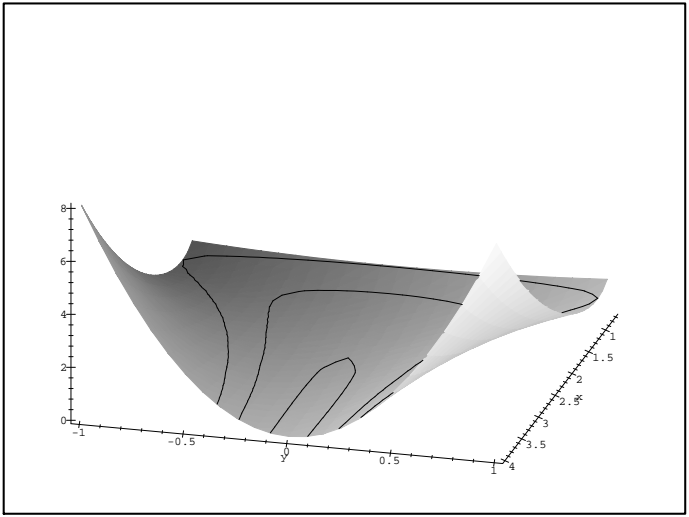, height=100mm, angle=-90}
\caption{Hyperbolic wall potential}
\label{figmodel}
\end{figure}

In figure (\ref{figmodel}) and (\ref{figfull}) we present the form of
the potential (\ref{lmodel}) and (\ref{L}), respectively and
draw corresponding equipotential lines.
Simple calculations show that the maximum of $g$ is reached at the
point $f=n/\sqrt{E}$.
The minimal accessible value of $f$ is $f_{min}=n/\sqrt{2E}$.
So, we deal with the potential for which equipotential lines go to infinity
along the $f$-axis.
Maximal value of $g/f$ in terms of $E, n$ and $\lambda$ is:
\be
\zeta\equiv\left(\frac gf\right)_{max}
=\frac{4\sqrt{2}}{3\sqrt{3}}\frac{E^{3/2}}{n^2\sqrt{\lambda}}
\label{F}
\ee
The condition $\zeta\ll 1$ guaranties that the system is always in the asymptotic
 region $g\sim \xi  f$ with $\xi \ll 1$.

We integrate this system asymptotically by using Bogolyubov-Krylov method
\cite {Bog}. According to this method one has to take
\be
g(t)=\sqrt{\frac{2\alpha}{\lambda f(t)}}\cos \varphi (t)
\ee
with some constant $\alpha$, then integrate the equation
\be
\frac{d^2f}{dt^2}+\alpha-\frac{n^2}{f^3}=0\label{fms}
\ee
Equation (\ref{fms}) can be easily integrated,
\begin{eqnarray}
t&=&\int\frac{df}{\sqrt{2\alpha(f_{max}-f)-\frac{n^2}{f^2f_{max}^2}(f_{max}^2-f^2)}}\\
\varphi&=&\sqrt{\lambda}\int f(t)dt+\varphi_0
\end{eqnarray}
Here constant of integration is chosen so that at $t=0$ the $f$-coordinate
takes its maximal value.
Energy $E$ and maximal deviation $f_{max}$ are related as follows:
\be
E=\alpha f_{max}+\frac{n^2}{2f_{max}^2}
\ee
One can invert this algebraic equation and get $f_{max}$ as function of dynamical
variables of our system (\ref{lmodel}).

As result we suggest that in close analogy with previous model
for $\zeta\ll 1$ the system becomes integrable and
$f_{max}$ is integral of motion. We shall justify this conjecture numerically.

\section{Numeric calculations.}
In this section we investigate the existence of chaos-order transition
for the hyperbolic model with reflecting wall, model with hyperbolic wall
(\ref{lmodel}) and system (\ref{L}) numerically.

At first we test the Poincar\`e sections for the model with a reflecting
wall.  The conservation of energy restricts any trajectory of the
four-dimensional phase space to a three-dimensional energy shell.  At a
given energy any additional constraint defines a two-dimensional surface
in the phase space, which is called the Poincar\`e section
It is convenient to take a constraint $g=0$.
All crossections of a trajectory with the surface are marked by points
on the $(f, p_f)$-plane.
On each figure we plot a Poincar\`e section for a set of trajectories to
show that behavior of the system does
not depend on the initial data. Any trajectory was integrated as long as the program
guarantees that the deviation of the energy is less then $0.1\%$.
Different colors correspond to different
trajectories with fixed parameters and random initial data.
Chaotic motion is characterized by
a set of randomly distributed points. Regular trajectories are
depicted by dotted curves.

In Figures (\ref{fig1}), (\ref{fig2}) and (\ref{fig3}) we plot Poincar\`e sections
for the different values of the reflecting wall coordinate
$l$ and with the same energy $E=1$.
The pictures show that for small values of the parameter $l$ chaotic
region is located near the wall. For large values of the parameter $l$
the points arrange into closed dotted curves.

An important characteristic of a dynamical system is the Lyapunov
exponent $\eta (t)$.
It has a positive limit: $(\lim_{t\to\infty}\eta>0)$
for a chaotic system and zero limit: $(\lim_{t\to\infty}\eta=0)$
for a regular one.
The calculations for the model with the hyperbolic wall (\ref{lmodel})
were performed for different values of $\zeta$ (\ref{F}). By changing the energy
$E$ with $\lambda$ and $n$ being fixed. We vary the parameter $\zeta$.
The program starts with random initial data with given
energy.  The program calculates the coordinates $f$ and $g$, the energy,
$f_{max}$ and the Lyapunov exponent.  Typical results for $\zeta= 1$
and $\zeta\ll 1$ are shown on Figures (\ref{fig4}) and (\ref{fig5})
respectively.  One can see that for $\zeta =1$ (Fig. \ref{fig4}, white
curve) Lyapunov exponent has a positive limit.  For $\zeta\ll 1$ the
Lyapunov exponent goes to zero (Fig. \ref{fig5}, white curve).  Parameter
$f_{max}$ (blue curve) for small $\zeta$ does not change with the time.
Energy (red line) shows that program works perfectly and energy
does conserve through the whole calculation time.
The numerical calculations show that for
$\zeta= 1$ the system (\ref{lmodel}) is stochastic and for
$\zeta\ll 1$ it is the integrable one, that confirm the analytical results of
sect. 4.

\begin{figure}
\epsfig{file=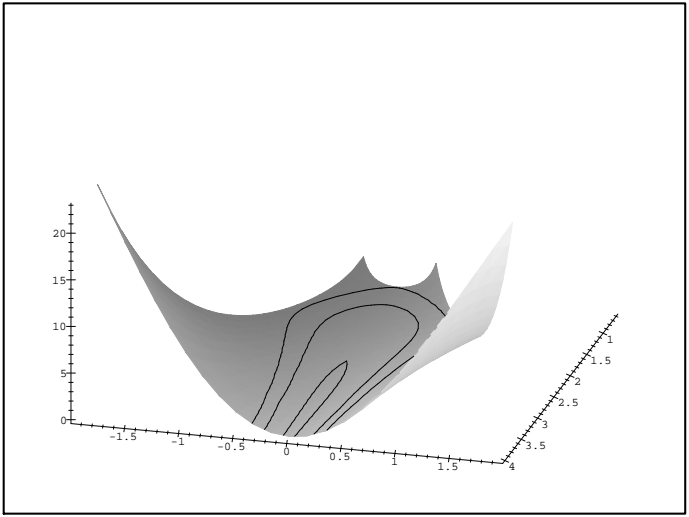, height=100mm, angle=-90}
\caption{Full potential}
\label{figfull}
\end{figure}

\begin{figure}
\epsfig{file=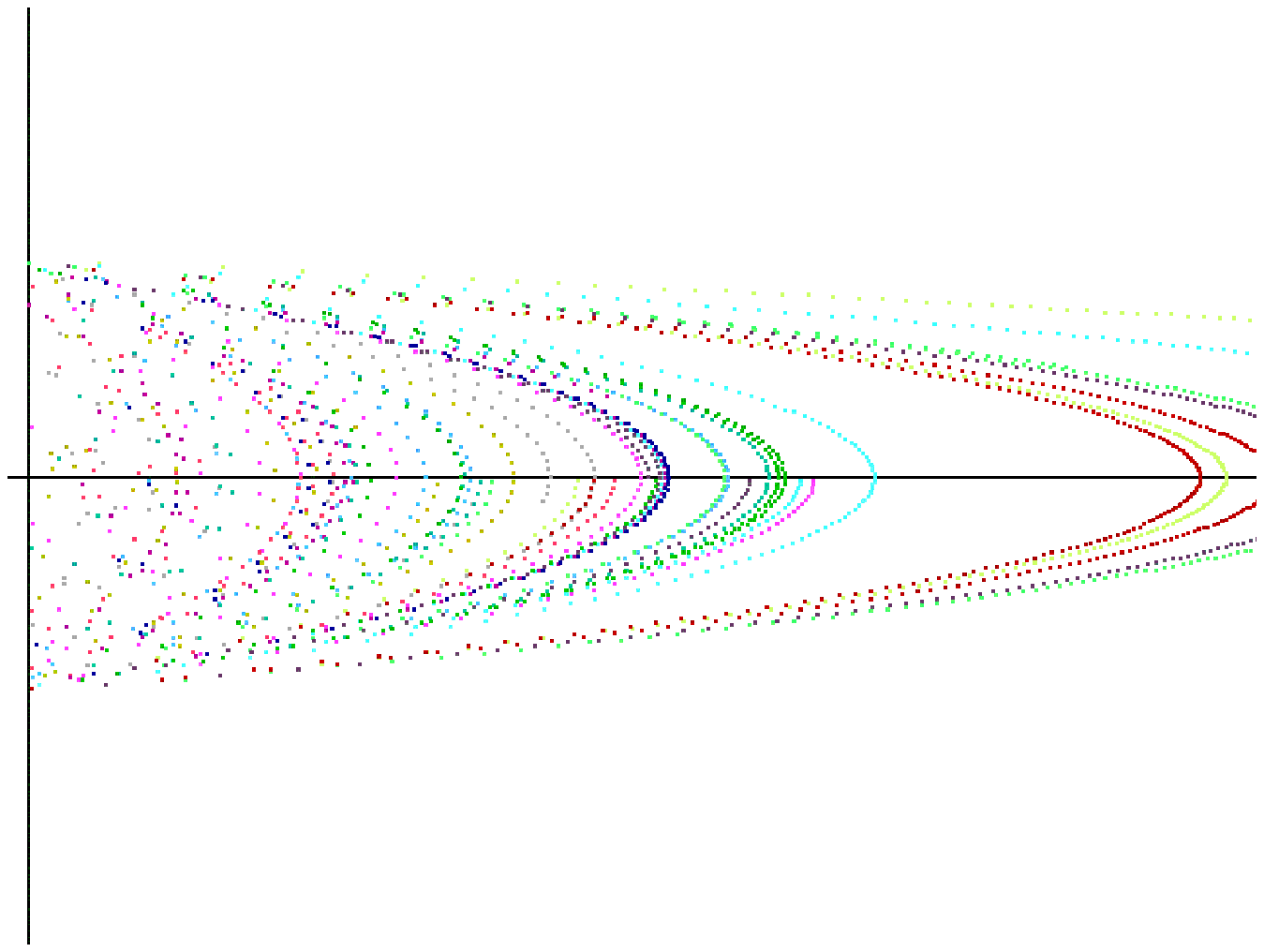}
\caption{Reflecting wall potential, $l=1,~E=1$}
\label{fig1}
\end{figure}

\begin{figure}
\epsfig{file=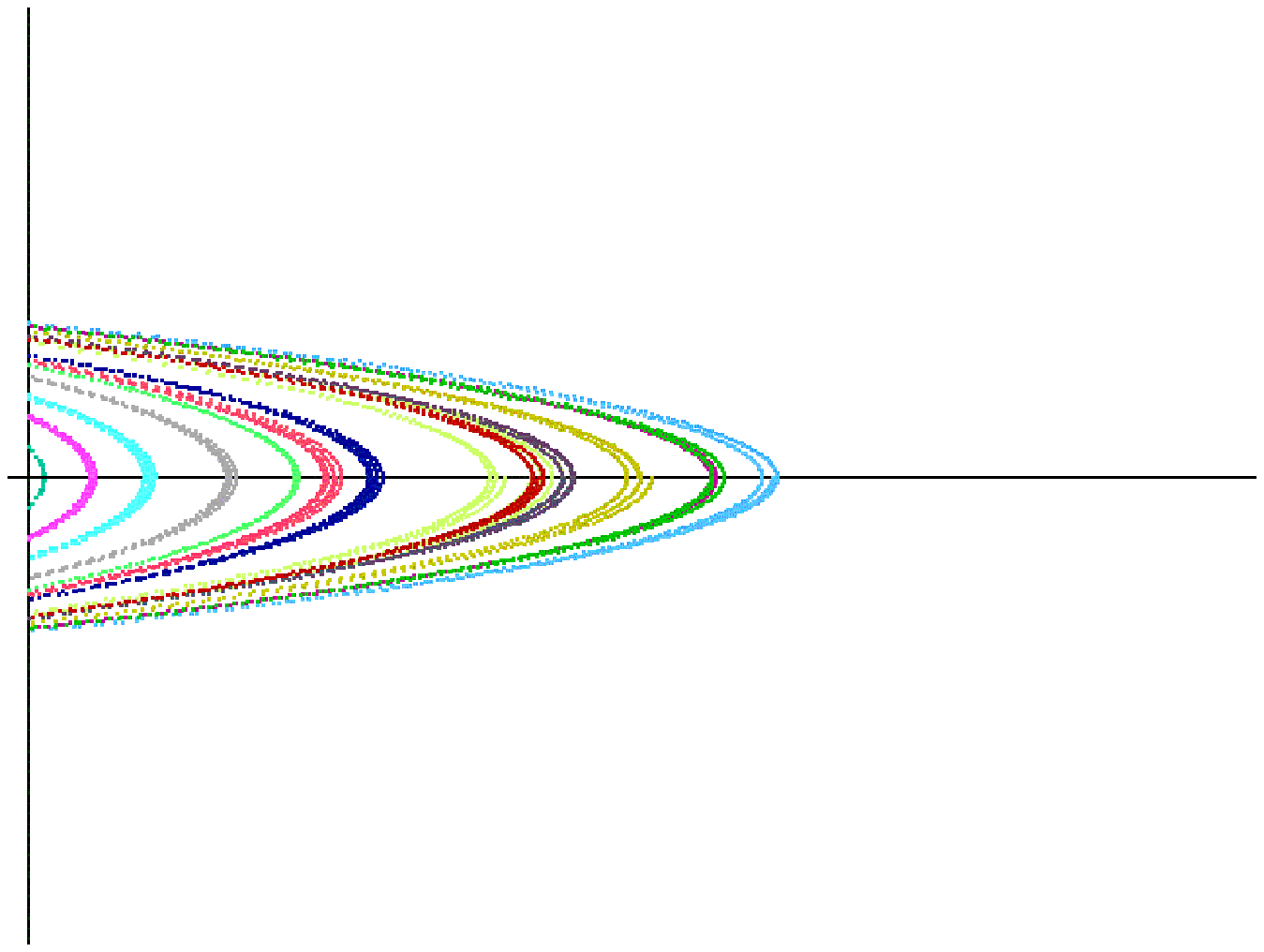}
\caption{Reflecting wall potential, $l=10,~E=1$}
\label{fig2}
\end{figure}

\begin{figure}
\epsfig{file=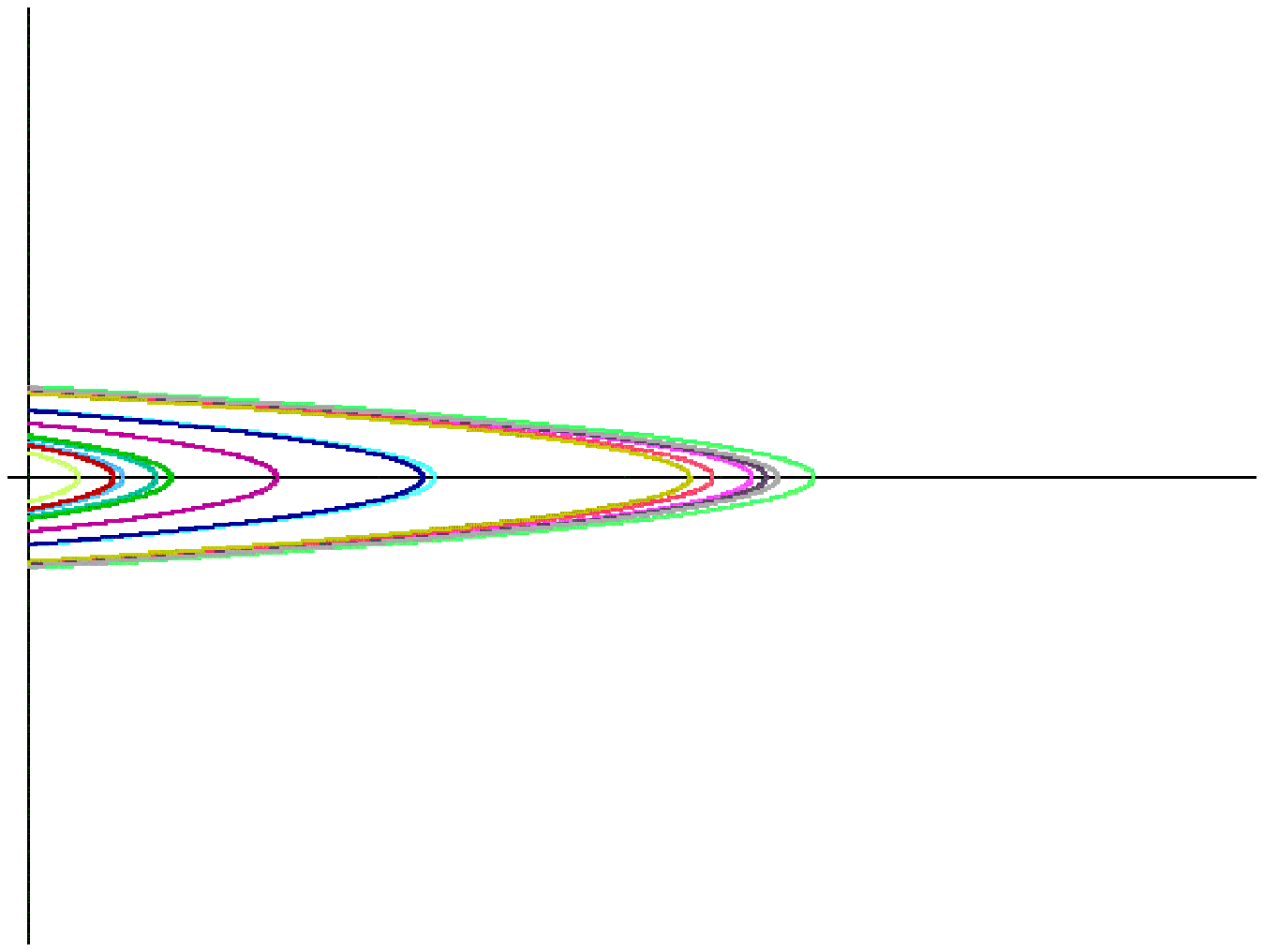}
\caption{Reflecting wall potential, $l=50,~E=1$}
\label{fig3}
\end{figure}

\begin{figure}
\epsfig{file=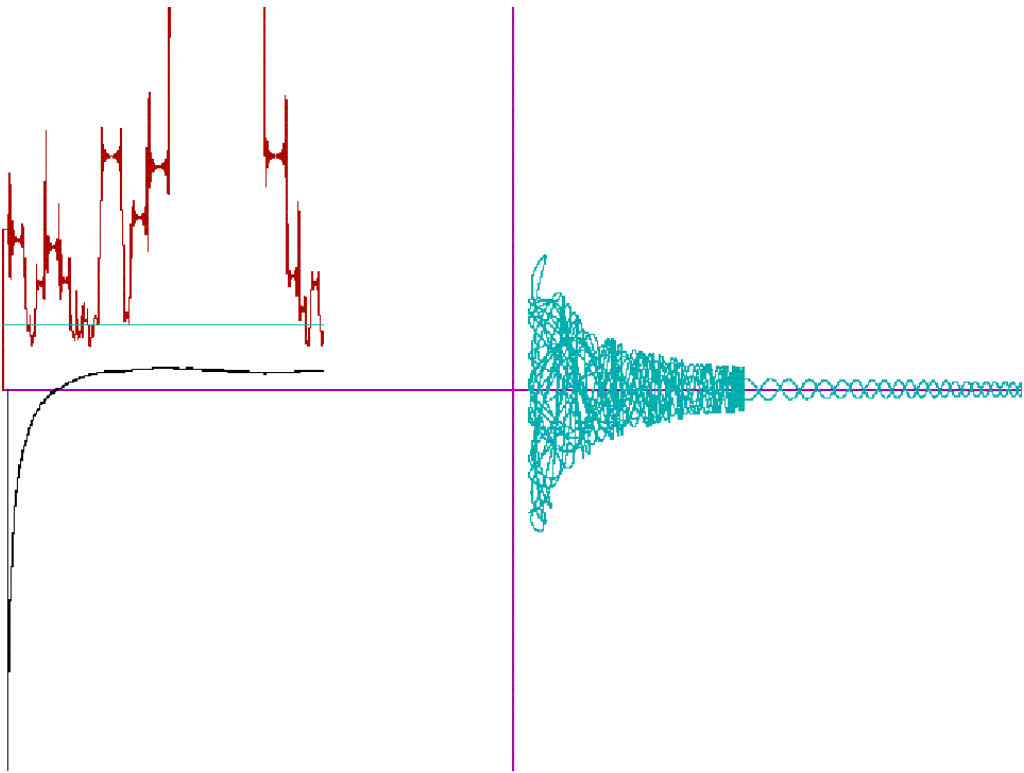}
\caption{Chaotic phase}
\label{fig4}
\end{figure}

\begin{figure}
\epsfig{file=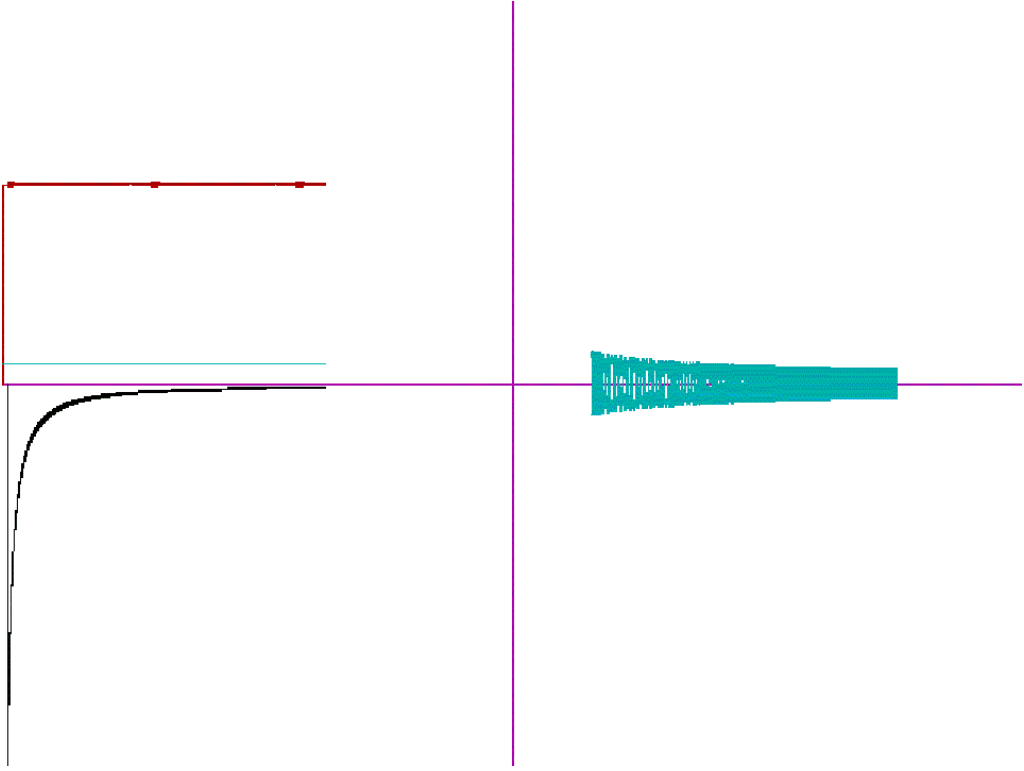}
\caption{Regular phase}
\label{fig5}
\end{figure}

\begin{figure}
\epsfig{file=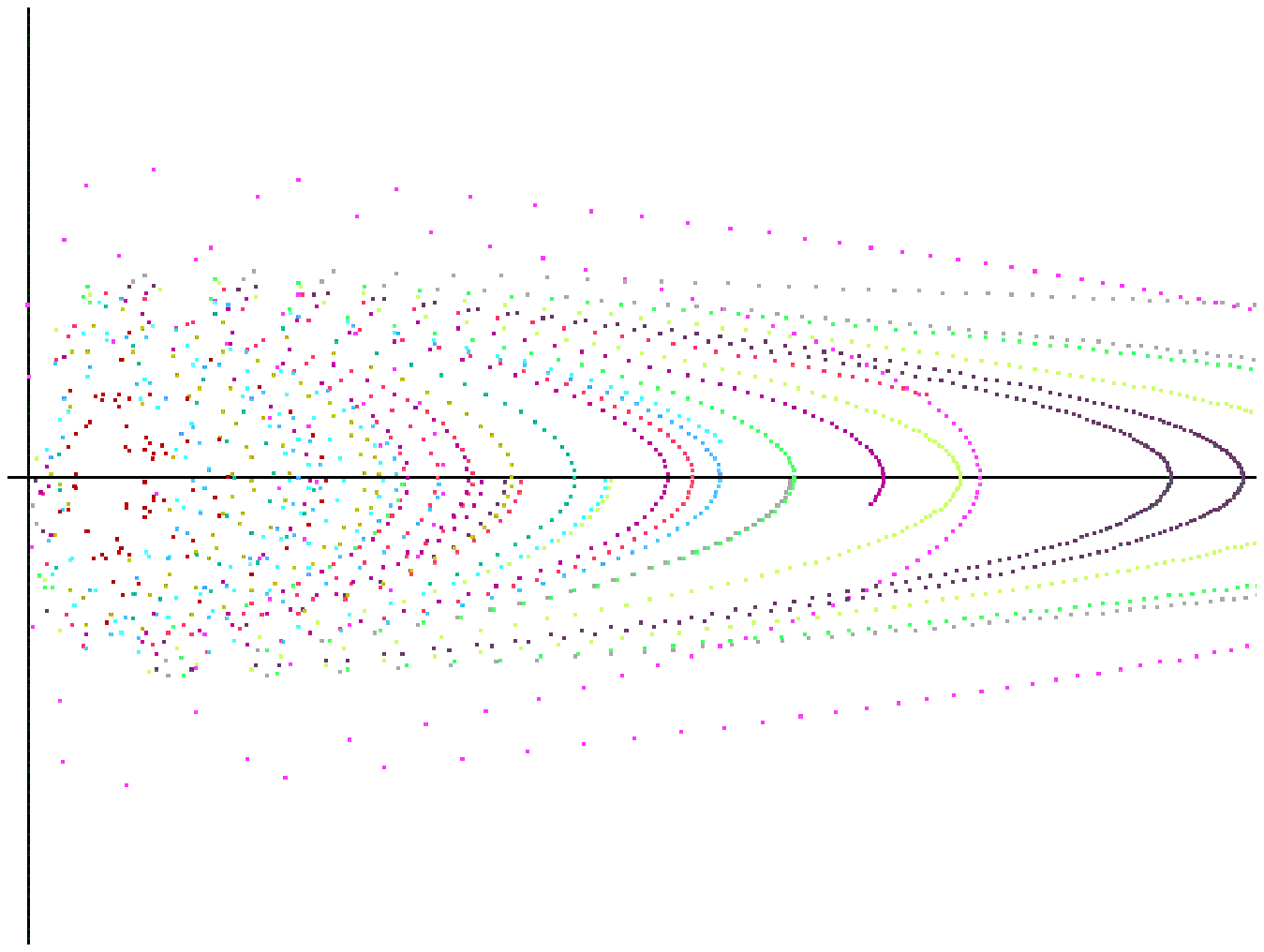}
\caption{Full potential, $n=1,~e=1$}
\label{fig6}
\end{figure}

\begin{figure}
\epsfig{file=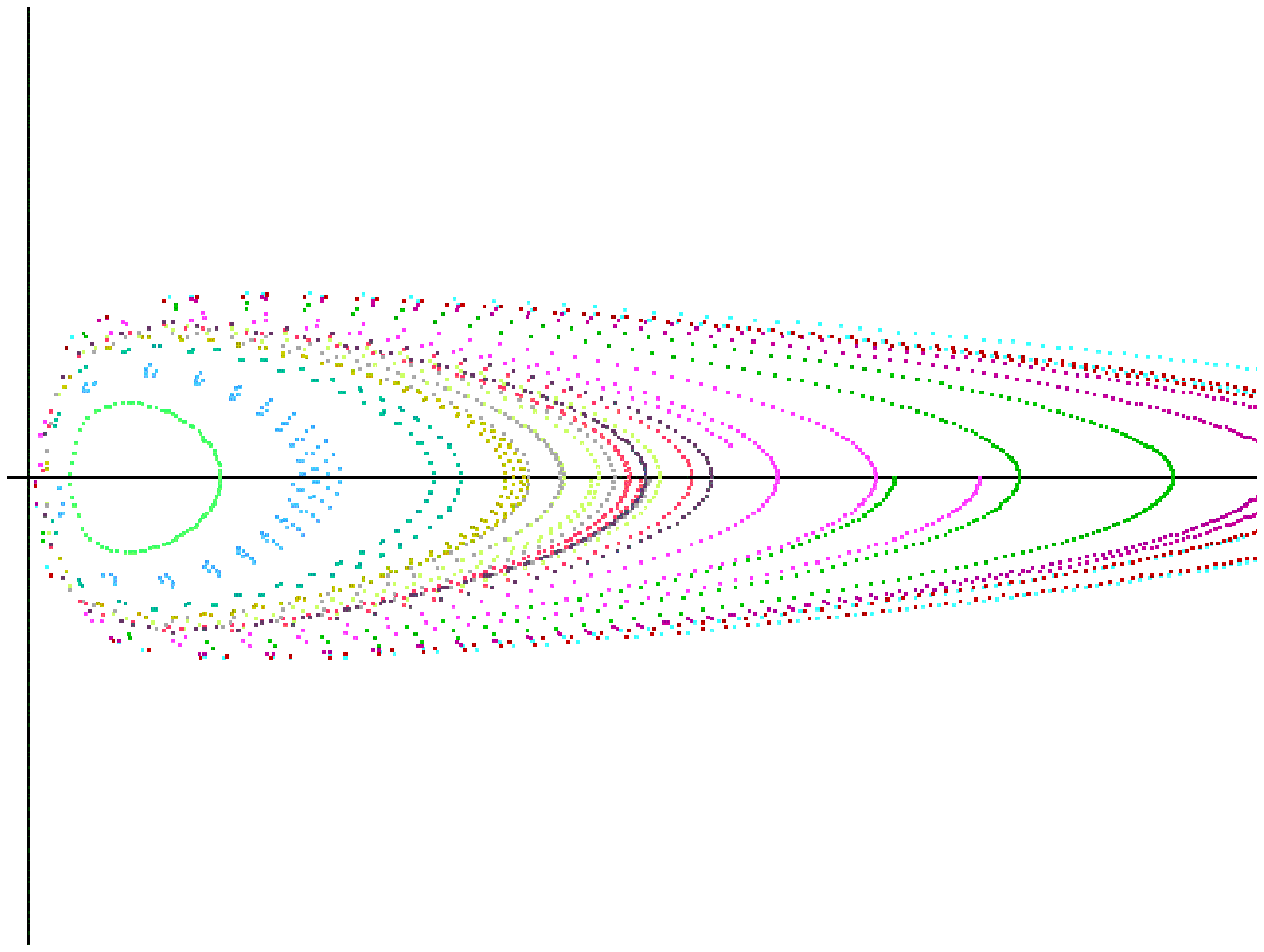}
\caption{Full potential, $n=2,~e=1$}
\label{fig7}
\end{figure}

\begin{figure}
\epsfig{file=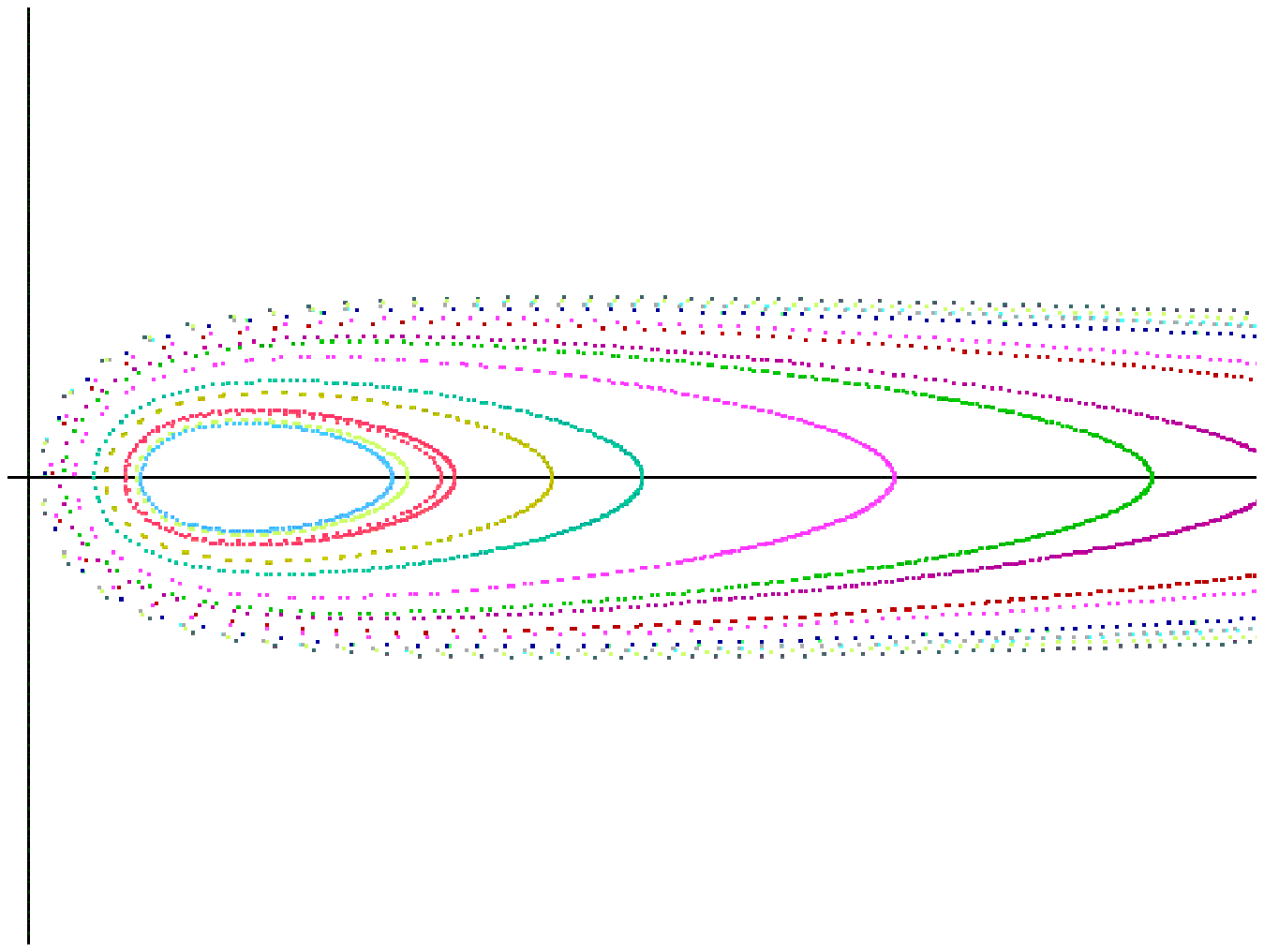}
\caption{Full potential, $n=5,~e=1$}
\label{fig8}
\end{figure}

Now we turn to the main model (\ref{L}). The dynamics of (\ref{L}) was
analyzed in \cite {AMRV} for relatively small (in units of $E$) values of
$n$ and we have found it to be fully chaotic. For the relatively large
values of $n$ the motion is confined into the region $\zeta\ll 1$ (see
Fig. \ref{figfull}).
In this region the model with hyperbolic wall (\ref{lmodel}) gives a good
approximation to the main model (\ref{L}). We conjecture the main model also
to be integrable for the large $n$.  In favor of this conjecture
in Figures (\ref{fig6}), (\ref{fig7})
and (\ref{fig8}) we plot the Poincar\`e sections for (\ref{L})
with different values of parameter $n$ and fixed energy $E=1$. For large
values of parameter $n$ there are only regular closed orbits.
We also test the Lyapunov exponent and obtained the pictures quite
similar to the ones for the model (\ref{lmodel}). Therefore we see that the
system exhibit the chaos-order transition governed by the characteristic
parameter $\zeta$.

$$~$$
{\bf ACKNOWLEDGMENT}
$$~$$
The authors are grateful to B.V.Medvedev for stimulating discussions.
I.A., P.M. and O.R.  are supported
in part by  RFFI grant 96-01-00608.
I.V. is supported  in part by RFFI grant 96-01-00312.
I.A. is supported in part by INTAS 96-0698.
$$~$$

\newpage

\newpage

\newpage

\newpage

\newpage

\end{document}